\documentclass{elsarticle}
\usepackage{mathrsfs}
\usepackage{amsmath,amssymb}
\usepackage{latexsym}
\usepackage{amsthm,a4wide}
\usepackage{graphicx}
\usepackage{CJK,CJKnumb}
\usepackage{bm}
\usepackage{subfigure}
\usepackage{caption}

\begin{document}

\graphicspath{{Graphics/}}

\title{A Simple Method to Construct Local Equilibrium Function for One Dimensional Lattice Boltzmann Method}
\tnotetext[t1]{This
work is partially supported by the Natural Science Foundation of
China(11071175).}

\author[rvt]{Peng Wang\corref{cor1}}
\ead{7790326@qq.com}
\author[rvt]{Shiqing Zhang}

\cortext[cor1]{Corresponding author: Tel.:+86 18602803561;fax:+86 0379 63316896.}
\address[rvt]{Yangtze Center of Mathematics and College of Mathematics, Sichuan University, Chengdu 610064, People's Republic of China}

\begin{abstract}

We have developed a simple method to construct local equilibrium function for one dimensional lattice Boltzmann method (LBM). This new method can make LBM model satisfy compressible flow with a flexible specific-heat ratio. Test cases, including the one dimensional Sod flow and one dimensional Lax flow are presented. Favorable results are obtained using proposed new method, indicating that the proposed method is potentially capable of constructing of the local equilibrium function for one dimensional LBM.
\end{abstract}
\begin{keyword}
Fluid mechanics \sep Lattice Boltzmann method \sep Local equilibrium function.
\end{keyword}
\maketitle

\section{Introduction}

The lattice Boltzmann method is a mesoscopic-based approach for solving the fluid flow problems at the macroscopic scales. In the LBM, the key issue is the determination
of the relaxation time $\tau$ and local equilibrium distribution function
$f^{(eq)}$. $\tau$ represents the relaxation time that distribution function $f$ relaxes to the local equilibrium function $f^{(eq)}$. By Chapman-Enskog expansion, relaxation time $\tau$ is linked to the viscosity and the pressure that satisfy the Navier-Stokes equations. $f^{(eq)}$ is local equilibrium function that is constrained by macroscopic conditions. What kind of flow that can be simulated is determined by the form of the $f^{(eq)}$. In this paper we mainly discuss a simple method to construct local equilibrium function $f^{(eq)}$.

In 1993, F. J. Alexander, S. Chen and J. D. Sterling (\cite{PhysRevE.47.R2249}) firstly to proposed a $f^{(eq)}$ for compressible viscosity flow. In 1994, Chen, etc. (\cite{PhysRevE.50.2776}) developed Alexander's $f^{(eq)}$ and obtained a model without nonlinearity deviations for recovering Navier-Stokes equations. In 2003, Watari and Tsuhara (\cite{PhysRevE.67.036306}) pointed out that Chen's model can only be deployed for small viscosity and temperature range. They introduced a global coefficients to construct $f^{(eq)}$ that make the model stable in a larger viscosity and temperature range. For the global coefficients in the Watari's model, more discrete velocity is required to make the $f^{(eq)}$ get correct form. 2007, Qu, etc. (\cite{PhysRevE.75.036706}) proposed a circular local equilibrium function. In Qu's model, the circular local equilibrium function cannot satisfy high order macroscopic statistic conditions that make the model can only recover to Euler equations. 2008, Li, etc. (\cite{S0129183108013126}) developed Qu's circular equilibrium function. In Li's model, the $f^{(eq)}$ satisfy all the macroscopic conditions which make correctly recovering of Navier-Stokes equations is possible. In the development history of $f^{(eq)}$ in LBM, we can find that the macroscopic conditions are crucial for the constructing of the $f^{(eq)}$.

In this work, based on the idea of assignment matrix to assign macroscopic conditions, we obtain a simple method to construct a local equilibrium distribution function for
compressible flows. In our
method, we do not consider the continuous form of local equilibrium function, and the discrete form of $f^{(eq)}$ is directly derived by assigning the macroscopic conditions, so that $f_i^{(eq)}$
do not contain free parameters. For numerical simulation, the differential form of the lattice-Boltzmann equation is
solved by the TVD scheme (\cite{ARTICLEZhangHanxinNON-OSCILLATORY}). Flows with
weak and strong shock waves were simulated successfully by the present model.
%

The paper is organized as follows. In Section \ref{BoltzmannBGK}, we introduce basic macroscopic conditions of LBM. In Section \ref{ConstructionLocalEquilibriumFunction}, the detail of our method to construct local equilibrium function is described. Firstly, we introduce assignment matrix to assign the non-energy conditions to construct $f_i^{(eq)}$, then, we use fixed specific rest energy to make the $f_i^{(eq)}$ satisfy energy conditions, Section \ref{NumericalExample} presents numerical results and discussions. In the last Section, we make some conclusions.




\section{Boltzmann-BGK equation for compressible flow and macroscopic  conditions}\label{BoltzmannBGK}
The standard dynamical theory of mesoscopic model
is described by the Boltzmann equation as following
\begin{equation}\label{Boltzmann_equation}
\frac{\partial f}{\partial t} + \bm{\xi} \cdot \nabla f = J,
\end{equation}
where $f$ is the distribution function (\cite{cercignani1988boltzmann}). The $\bm{\xi}$
is the particle velocity vector, and $J$ represents the
collision term, we can simplify the collision term $J$ to the
following formulation (\cite{PhysRev.94.511}):
\begin{equation}\label{continuous_bgk_equation}
\frac{\partial f}{\partial t} + \bm{\xi} \cdot \nabla f =
\frac{1}{\tau}(f^{(eq)} - f  ),
\end{equation}
where $\tau$ is the relaxation time and $f^{(eq)}$ is the
local equilibrium distribution function.

The moments of $f^{(eq)}$ are
\begin{subequations}
\begin{equation} \label{ContinuousStatisticalVariables1}
\int f^{(eq)} d\bm{\xi} = \rho,
\end{equation}
\begin{equation}\label{ContinuousStatisticalVariables2}
\int f^{(eq)}\xi_\alpha d\bm{\xi}  = \rho u_{\alpha},
\end{equation}
\begin{equation}\label{ContinuousStatisticalVariables3}
\int f^{(eq)}\xi_\alpha \xi_\beta d\bm{\xi} =  \rho u_{\alpha} u_{\beta} + p \delta_{\alpha\beta},
\end{equation}
\begin{equation}\label{ContinuousStatisticalVariables3}
\int f^{(eq)}\xi_\alpha \xi_\beta \xi_\gamma d\bm{\xi}  =  \rho u_\alpha u_\beta u_\gamma + p( u_{\gamma} \delta_{\alpha \beta} + u_{\beta}\delta_{\alpha\gamma} + u_{\alpha} \delta_{\beta\gamma}  ),
\end{equation}
\begin{equation}\label{ContinuousStatisticalVariables5}
\int f^{(eq)}(\frac{1}{2}|\bm{\xi}|^2 + \zeta) d\bm{\xi} = \rho E ,
\end{equation}
\begin{equation}\label{ContinuousStatisticalVariables6}
\int f^{(eq)}(\frac{1}{2}|\bm{\xi}|^2 + \zeta)\xi_\alpha d\bm{\xi} = (\rho E + p)u_{\alpha},
\end{equation}
\begin{equation}\label{ContinuousStatisticalVariablesLast}
\begin{aligned}
\int f^{(eq)}(\frac{1}{2}|\bm{\xi}|^2 + \zeta)\xi_\alpha \xi_\beta d\bm{\xi}  =(\rho E + 2 p )u_{\alpha}u_{\beta} + p(E + R_g T)\delta_{\alpha\beta},
\end{aligned}
\end{equation}
\end{subequations}
where the $\zeta$ is the specific rest energy (\cite{PhysRevE.58.7283}, \cite{PhysRevE.75.036706}) and $R_g$ is gas constant. Consider the connection between mesoscopic variable and macroscopic variable:
\begin{gather}
\frac{1}{2} \rho c^2 = p,\\
\frac{D}{4} \rho c^2 + \rho \zeta = \rho \varepsilon,
\end{gather}
where $D$ is the spatial dimension, $c$ is peculiar speed and $\varepsilon$ is the internal energy. For the polytropic gas, we have
\begin{equation}\label{SpecificRestEnergy}
\zeta = [1-\frac{D}{2}(\gamma - 1)]\varepsilon,\\
\end{equation}
\begin{equation}
c = \sqrt{2(\gamma - 1)\varepsilon},
\end{equation}
where $\gamma$ is the specific heat capacity ratio.

Applying Chapman-Enskog expansion (\cite{BOOKHeYL}) and moments (\ref{ContinuousStatisticalVariables1}) - (\ref{ContinuousStatisticalVariablesLast}) to equations (\ref{continuous_bgk_equation}), we can recover the macroscopic Navier-Stokes equations without body force:

\begin{equation*}
\begin{aligned}
& \frac{\partial \rho}{\partial t} + \nabla \cdot (\rho \bm{u}) = 0,\\
&\frac{\partial (\rho \bm{u})}{\partial t} +\nabla\cdot (\rho\bm{u}\otimes\bm{u}) =
-\nabla p + \mu \Delta \bm{u} + (\mu + \mu')\nabla(\nabla\cdot\bm{u}),\\
&\frac{\partial (\rho E )}{\partial t} +\nabla\cdot [(\rho E + p)\bm{u}] =
\nabla \cdot (\lambda \nabla T)  + \nabla \cdot (\bm{u} \cdot \Pi),
\end{aligned}
\end{equation*}
where $\Pi = \mu[\nabla \bm{u} + (\nabla \bm{u})^T - \frac{2}{D} (\nabla \cdot \bm{u})I]$, $\mu$ is viscosity, $\mu'$ is the bulk viscosity and $\lambda$ is thermal conductivity. We have
\begin{equation}
\mu = p \tau,\quad \lambda = (1+ \frac{b}{2})R_g p \tau.
\end{equation}

\section{Construction of discretization local equilibrium function}\label{ConstructionLocalEquilibriumFunction}

In above section, we known that the continuous $f^{(eq)}$ satisfy the macroscopic  condition (\ref{ContinuousStatisticalVariables1}) - (\ref{ContinuousStatisticalVariablesLast}). For the lattice Boltzmann method, we need to known the discrete $f_i^{(eq)}$ that deploy to the corresponding lattice microscopic velocity. The moments of discrete $f_i^{(eq)}$ are
\begin{subequations}
\begin{equation} \label{StatisticalVariables1}
 \sum_{i} f_{i}^{(eq)} = \rho,
\end{equation}
\begin{equation}\label{StatisticalVariables2}
 \sum_{i} f_{i}^{(eq)} \xi_{i\alpha}  = \rho u_{\alpha},
\end{equation}
\begin{equation}\label{StatisticalVariables3}
 \sum_{i} f_{i}^{(eq)} \xi_{i\alpha} \xi_{i\beta} =  \rho u_{\alpha} u_{\beta} + p \delta_{\alpha\beta},
\end{equation}
\begin{equation}\label{StatisticalVariables4}
 \sum_{i} f_{i}^{(eq)} \xi_{i\alpha} \xi_{i\beta} \xi_{i\gamma} =  \rho u_\alpha u_\beta u_\gamma + p( u_{\gamma} \delta_{\alpha \beta} + u_{\beta}\delta_{\alpha\gamma} + u_{\alpha} \delta_{\beta\gamma}  ),
\end{equation}
\begin{equation}\label{StatisticalEnergyVariables1}
 \sum_{i} f_{i}^{(eq)} (\frac{1}{2}|\bm{\xi}_{i}|^2 + \zeta) = \rho E ,
\end{equation}
\begin{equation}\label{StatisticalEnergyVariables2}
 \sum_{i} f_{i}^{(eq)}(\frac{1}{2}|\bm{\xi}_{i}|^2 + \zeta) \xi_{i\alpha} = (\rho E + p)u_{\alpha},
\end{equation}
\begin{equation}\label{StatisticalVariablesLast}
\begin{aligned}
 \sum_{i} f_{i}^{(eq)}(\frac{1}{2}|\bm{\xi}_{i}|^2 + \zeta)\xi_{i\alpha} \xi_{i\beta}  =(\rho E + 2 p )u_{\alpha}u_{\beta} + p(E + R_g T)\delta_{\alpha\beta}.
\end{aligned}
\end{equation}
\end{subequations}

By equations (\ref{StatisticalVariables1}) - (\ref{StatisticalVariablesLast}), the macroscopic conditions can be classified as non-energy statistic conditions and energy statistic conditions. The non-energy statistic conditions are
\begin{equation}
S_N = \left(
  \begin{array}{c}
  \rho \\
  \rho u_{\alpha}\\
  \rho u_{\beta}\\
  \rho u_{\alpha}u_{\beta}\\
   \rho u_{\alpha}u_{\alpha}+p\\
    \vdots\\
  \rho u_{\alpha}u_{\beta}u_{\gamma}+ p( u_{\gamma} \delta_{\alpha \beta} + u_{\beta}\delta_{\alpha\gamma} + u_{\alpha} \delta_{\beta\gamma}  )
  \end{array}\right).
\end{equation}
From equation (\ref{StatisticalVariables1}) - (\ref{StatisticalVariablesLast}), energy statistic conditions are translations of non-energy statistic conditions at some energy level. Therefore, $f_{i}^{(eq)}$ should only conserve the non-energy statistic conditions.
For the number of non-energy statistic conditions that we need to conserve, we choose corresponding number of microscopic $\xi_i$, i.e., the number of lattice microscopic velocity $\xi_i$ is the same as the  non-energy statistic conditions.  The square matrix that made up of moments of microscopic velocity is
\begin{equation}
V  = \left(
  \begin{array}{cccc}
  1 & 1&  \cdots & 1 \\
  \xi_{1\alpha} &  \xi_{2\alpha}  & \cdots &  \xi_{i\alpha} \\
  \xi_{1\beta} &  \xi_{2\beta}  & \cdots &  \xi_{i\beta} \\
  \vdots        &\vdots                  & \cdots & \vdots    \\
   \xi_{1\alpha}\xi_{1\beta} &  \xi_{2\alpha}\xi_{2\beta}  & \cdots &  \xi_{i\alpha}\xi_{i\beta} \\
   \vdots        &\vdots                  & \cdots & \vdots    \\
   \xi_{1\alpha}\xi_{1\beta} \xi_{1\gamma} &  \xi_{2\alpha}\xi_{2\beta}\xi_{2\gamma} & \cdots &  \xi_{i\alpha}\xi_{i\beta}\xi_{i\gamma} \\
  \end{array}
\right).
\end{equation}
Define the $V^{-1}$ as the assign matrix, where the $V^{-1}$ is the inverse of $V$, we have discrete form $f_i^{(eq)}$ as
\begin{equation}\label{DiscreteLocalEquilibriumFunction}
\left(
\begin{array}{c}
f_{1}^{(eq)}\\
f_{2}^{(eq)}\\
\vdots\\
f_{i}^{(eq)}
\end{array}\right) = V^{-1} S_N.
\end{equation}
By equation (\ref{DiscreteLocalEquilibriumFunction}), we can simply find that the $V * V^{-1}S_N$ satisfy equation (\ref{StatisticalVariables1}), (\ref{StatisticalVariables2}), (\ref{StatisticalVariables3}) and (\ref{StatisticalVariables4}). However, we know the $f_i^{(eq)}$ satisfy non-energy conditions, we can not make the $f_i^{(eq)}$ satisfy the energy macroscopic  conditions
\[
 S_E = \left(
  \begin{array}{c}
  \rho {E} \\
  (\rho {E} + p)u_{\alpha}\\
\vdots\\
  (\rho {E} + 2 p )u_{\alpha}u_{\beta} + p({E} + R_g T)\delta_{\alpha\beta}\\
  \end{array}\right).
\]
For thermal flow, $\zeta$ is the specific rest energy. In equation (\ref{ContinuousStatisticalVariables5}), (\ref{ContinuousStatisticalVariables6}) and (\ref{ContinuousStatisticalVariablesLast}), $\zeta$ is a microscopic variable and $\zeta$ is unknown that make the total energy $E$ can not be calculated from the moments of the distribution function $f_i$. Therefore, the introduction of a fixed $\zeta_j$
to assign $f_i^{(eq)}$ is necessary. The introduction of fixed $\zeta_j$ can also solve the problem of $f_i^{(eq)}$ which do not satisfy energy conditions. By Qu's work (\cite{PhysRevE.75.036706}), linearly assigning $f_i^{(eq)}$ onto two energy levels $\zeta_1$ and $\zeta_2$, $\zeta_1 = 0$ and $\zeta_2 > e_{max}$ are the minimum and maximum stagnation energy in the whole flow field. We have
\begin{equation*}
\begin{aligned}
f_{i,\zeta_1}^{(eq)} = f_{i}^{(eq)} \frac{\zeta_2 - \zeta }{\zeta_2},\\
f_{i,\zeta_2}^{(eq)} = f_{i}^{(eq)} \frac{\zeta }{\zeta_2}.\\
\end{aligned}
\end{equation*}

By equation (\ref{StatisticalVariables1}) - (\ref{StatisticalVariablesLast}), (\ref{SpecificRestEnergy}) and (\ref{DiscreteLocalEquilibriumFunction}), we can easily verify that $f_{i,\zeta_j}^{(eq)}$ satisfy
\begin{subequations}
\begin{equation} \label{zetaStatisticalVariables1}
 \sum_{j} \sum_{i} f_{i,\zeta_j}^{(eq)} = \rho,
\end{equation}
\begin{equation}\label{zetaStatisticalVariables2}
 \sum_{j} \sum_{i} f_{i,\zeta_j}^{(eq)}\xi_\alpha  = \rho u_{\alpha},
\end{equation}
\begin{equation}\label{zetaStatisticalVariables3}
 \sum_{j} \sum_{i} f_{i,\zeta_j}^{(eq)}\xi_\alpha \xi_\beta =  \rho u_{\alpha} u_{\beta} + p \delta_{\alpha\beta},
\end{equation}
\begin{equation}\label{zetaStatisticalVariables4}
 \sum_{j} \sum_{i} f_{i,\zeta_j}^{(eq)}\xi_\alpha \xi_\beta \xi_\gamma =  \rho u_\alpha u_\beta u_\gamma + p( u_{\gamma} \delta_{\alpha \beta} + u_{\beta}\delta_{\alpha\gamma} + u_{\alpha} \delta_{\beta\gamma}  ),
\end{equation}
\begin{equation}\label{zetaStatisticalEnergyVariables1}
 \sum_{j} \sum_{i} f_{i,\zeta_j}^{(eq)}(\frac{1}{2}|\bm{\xi}|^2 + \zeta_j) = \rho E ,
\end{equation}\label{zetaStatisticalEnergyVariables2}
\begin{equation}
 \sum_{j} \sum_{i} f_{i,\zeta_j}^{(eq)}(\frac{1}{2}|\bm{\xi}|^2 + \zeta_j)\xi_\alpha = (\rho E + p)u_{\alpha},
\end{equation}\label{zetaStatisticalEnergyVariables3}
\begin{equation}\label{zetaStatisticalVariablesLast}
\begin{aligned}
 \sum_{j} \sum_{i} f_{i,\zeta_j}^{(eq)}(\frac{1}{2}|\bm{\xi}|^2 + \zeta_j)\xi_\alpha \xi_\beta  =(\rho E + 2 p )u_{\alpha}u_{\beta} + p(E + R_g T)\delta_{\alpha\beta}.
\end{aligned}
\end{equation}
\end{subequations}

By this stage, entire $f_i^{(eq)}$ of the lattice Boltzmann method for compressible flow with a flexible specific-heat ratio is established.

\section{Calculation scheme}
\subsection{Time discrete}
For time discrete, we calculates advection term $\vec{e}_i\cdot
\nabla f_i(t_\xi,\vec{x}) \Delta t$ as an explicit finite-difference
form and the collision term $ \frac{\Delta
t}{\tau}(f_i^{(eq)}(t_\xi,\vec{x}) -
    f_i(t_\xi,\vec{x}))$ as a
implicit finite-difference form, and we got
\begin{equation}\label{iterative_scheme}
f_i^{n+1} - f_i^n + \Delta t \vec{e}_i \cdot \nabla f_i^n = \Delta t
[\theta J_i^{n+1} + (1-\theta)J_i^n],
\end{equation}
where $x \leqslant t_\xi \leqslant x+\Delta x$, $J_i^n = \frac{\Delta
t}{\tau}(f_i^{(eq)}(t,\vec{x}) -
    f_i(t,\vec{x})),\, J_i^{n+1} = \frac{\Delta
t}{\tau}(f_i^{(eq)}(t+\Delta t,\vec{x}) -
    f_i(t+\Delta t,\vec{x})).$ and $\theta$ represents the degree of implicity, in this paper
we set $\theta$ to be $0.5$.

Z.L Guo and T.S Zhao (\cite{PhysRevE.67.066709}) introduced a new distribution function to remove
the implicity of the equation (\ref{iterative_scheme}), and the new distribution function is
\begin{equation}\label{new_df}
g_i = f_i +   \pi\theta(f_i - f_i^{(eq)}),
\end{equation}
where $\pi = \Delta t/\tau$. Applying this new distribution function
to equation (\ref{iterative_scheme}), we can vanish the implicity
of the collision term and got
\begin{equation}
\begin{aligned}
&g_i^{n+1} = -\Delta t \vec{e}_i\cdot \nabla f_i^n + (1-\pi + \pi
\theta)f_i^{n} + \pi(1-\theta)f_i^{(eq),n},\\
&f_i^{n+1} = \frac{1}{1+ \pi \theta} (g_i^{n+1} + \pi \theta f^{(eq),n+1}).
\end{aligned}
\end{equation}

\subsection{Space discrete}
For space discrete, we define
\begin{equation}
\begin{aligned}
f_i(I,J) = f_i(x_I,y_J),\quad f_i(I - 1,J) = f_i(x_I - \Delta x,y_J),
\end{aligned}
\end{equation}
where I and J are node indexes. The second-order TVD scheme is
\begin{equation}
\begin{aligned}
\frac{\partial (e_{i\alpha}f_i)}{\partial x} = \frac{1}{\Delta t} [ F_i(I+\frac{1}{2},J) - F_i(I-\frac{1}{2},J) ],
\end{aligned}
\end{equation}
where $F_i(I+\frac{1}{2},J)$ is the numerical
flux at the interface of $x_I + \Delta x/2$, and is defined
as:
\begin{equation}
\begin{aligned}
F_i(I+\frac{1}{2},J) = F^L_i(I+\frac{1}{2},J) + F^R_i(I+\frac{1}{2},J),
\end{aligned}
\end{equation}
where
\begin{subequations}
\begin{equation}\label{fluxlimiter1}
 F^L_i(I+\frac{1}{2},J) =  F^{+}_i(I,J) + \frac{1}{2} min\,mod \,{(\Delta F^{+}_i(I+\frac{1}{2},J), \Delta F^{+}_i(I-\frac{1}{2},J))},
\end{equation}
\begin{equation}\label{fluxlimiter2}
 F^R_i(I+\frac{1}{2},J) =  F^{-}_i(I+1,J) - \frac{1}{2} min\,mod \,{(\Delta F^{-}_i(I+\frac{1}{2},J), \Delta F^{-}_i(I+\frac{3}{2},J))},
\end{equation}
\begin{equation}
 F^{+}_i(I,J) = \frac{1}{2}(e_{i\alpha} + \left|e_{i\alpha}\right|)f_i(I,J),
 \end{equation}
 \begin{equation}
 F^{-}_i(I,J) = \frac{1}{2}(e_{i\alpha} - \left|e_{i\alpha}\right|)f_i(I,J),
\end{equation}
and
\begin{equation}
\Delta F^{\pm}_i(I+\frac{1}{2},J) =   F^{\pm}_i(I + 1,J) - F^{\pm}_i(I,J).
\end{equation}
\end{subequations}
The function $min\,mod\,(\,\,,\,\, )$ in (\ref{fluxlimiter1}) and (\ref{fluxlimiter2}) is the
flux limiter (\cite{ARTICLEZhangHanxinNON-OSCILLATORY}, \cite{Y.WANGarticleIMPLICIT}).


\section{Numerical examples}\label{NumericalExample}
In this section, we will applied new $f_i^{(eq)}$ to Sod shock tube flow and Lax shock tube
flow to test capability of
this new method.
In numerical computations, the dimensionless form is preferred.
There are three independent reference variables for
normalization, which are the reference density $\rho_0$, reference
length $L_0$, and reference internal energy $e_0$. With the three
reference variables, other reference variables and non-dimensional
variables can be defined as
\begin{equation}
\begin{aligned}
u_0 = \sqrt{e_0}, t_0 = \frac{L_0}{u_0},\hat{t} = \frac{t}{t_0}, \hat{x} = \frac{x}{L_0}, \hat{\rho} = \frac{\rho}{\rho_0},\hat{u} = \frac{u}{u_0}, \hat{e} = \frac{e}{e_0}.
\end{aligned}
\end{equation}

For one dimensional LBM, we choose discrete velocity as $e_1 = 1,e_2 = -1, e_3 = 2, e_4 = -2$. By the assignment matrix, we have
\begin{equation}
\begin{aligned}
f_{1}^{(eq)} &= - \mathrm{\rho}\, \left(\frac{c^2\, u_{x}}{4} + \frac{c^2}{12} + \frac{{u_{x}}^3}{6} + \frac{{u_{x}}^2}{6} - \frac{2\, u_{x}}{3} - \frac{2}{3}\right)\\
f_{2}^{(eq)} &= - \mathrm{\rho}\, \left( - \frac{c^2\, u_{x}}{4} + \frac{c^2}{12} - \frac{{u_{x}}^3}{6} + \frac{{u_{x}}^2}{6} + \frac{2\, u_{x}}{3} - \frac{2}{3}\right)\\
f_{3}^{(eq)} &= \mathrm{\rho}\, \left(\frac{c^2\, u_{x}}{8} + \frac{c^2}{12} + \frac{{u_{x}}^3}{12} + \frac{{u_{x}}^2}{6} - \frac{u_{x}}{12} - \frac{1}{6}\right)\\
f_{4}^{(eq)} &= \mathrm{\rho}\, \left( - \frac{c^2\, u_{x}}{8} + \frac{c^2}{12} - \frac{{u_{x}}^3}{12} + \frac{{u_{x}}^2}{6} + \frac{u_{x}}{12} - \frac{1}{6}\right)
\end{aligned}
\end{equation}

\subsection{Sod shock tube}
First, we use the Sod shock tube to test the proposed model. Sod shock tube is a famous unsteady flow, which includes shock wave. The initial
condition is given as
\begin{equation}
\begin{aligned}
(\hat{\rho}_L,\hat{u}_L,\hat{e}_L) &= (1,0,2.5)\quad (-0.5< \hat{x} < 0)\\
(\hat{\rho}_R,\hat{u}_R,\hat{e}_R) &= (0.125,0,2)\quad (0< \hat{x} < 0.5)
\end{aligned}
\end{equation}

\begin{figure}[!h]
\centering
\begin{minipage}[t]{0.45\linewidth}
\centering
\includegraphics[width=\linewidth]{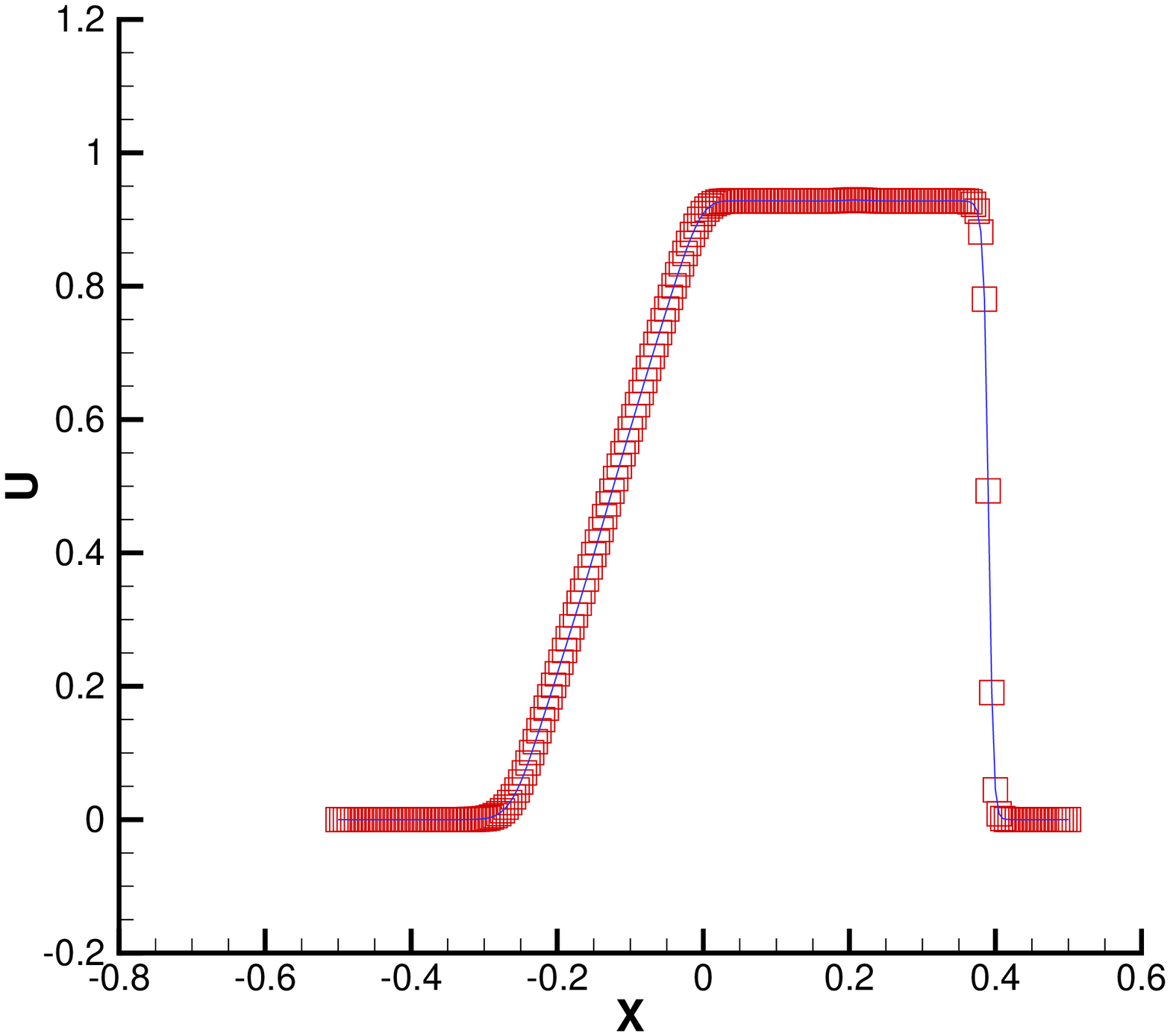}
\end{minipage}\quad
\begin{minipage}[t]{0.45\linewidth}
\centering
\includegraphics[width=\linewidth]{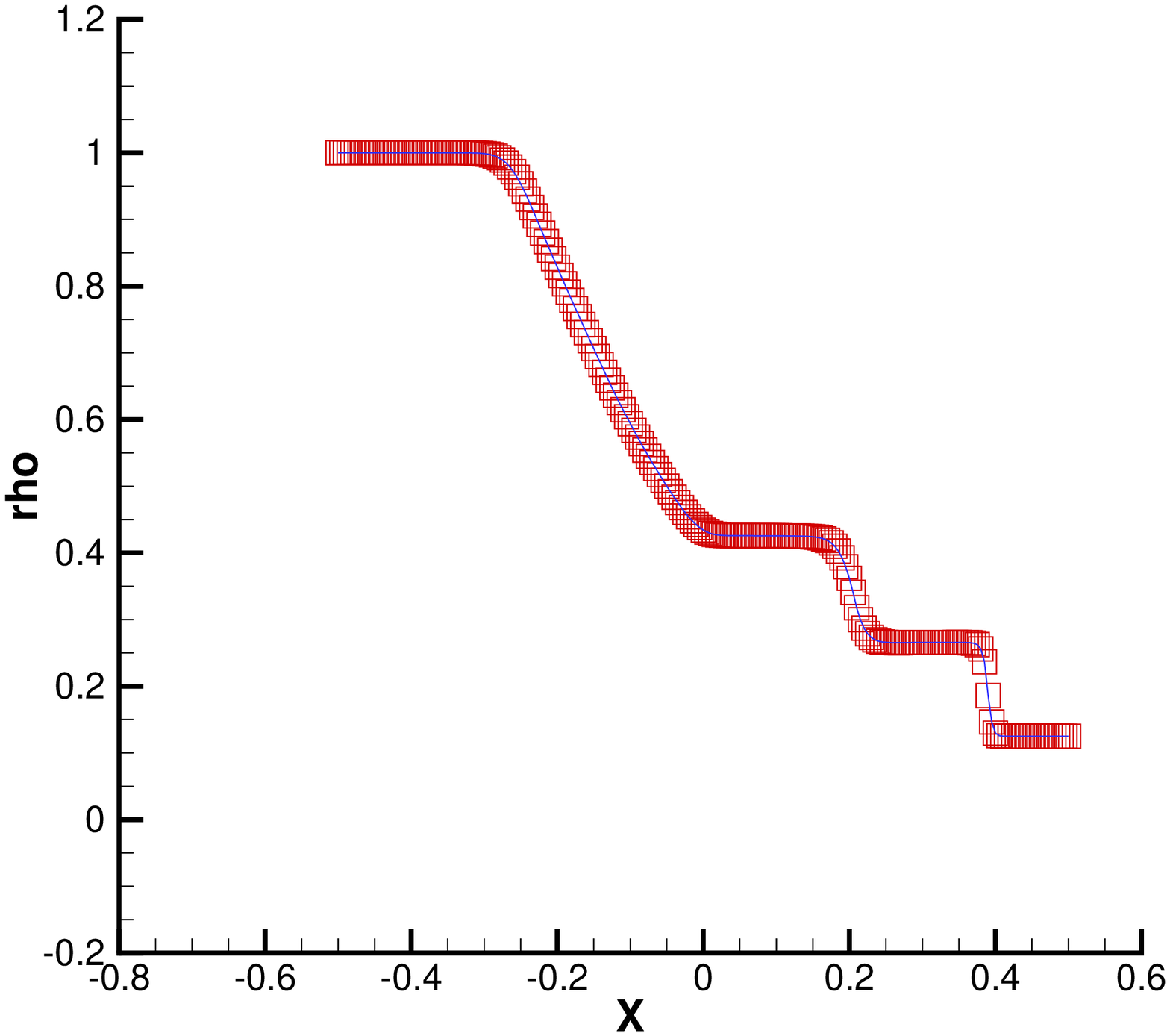}
\end{minipage}\\
\begin{minipage}[t]{0.45\linewidth}
\centering
\includegraphics[width=\linewidth]{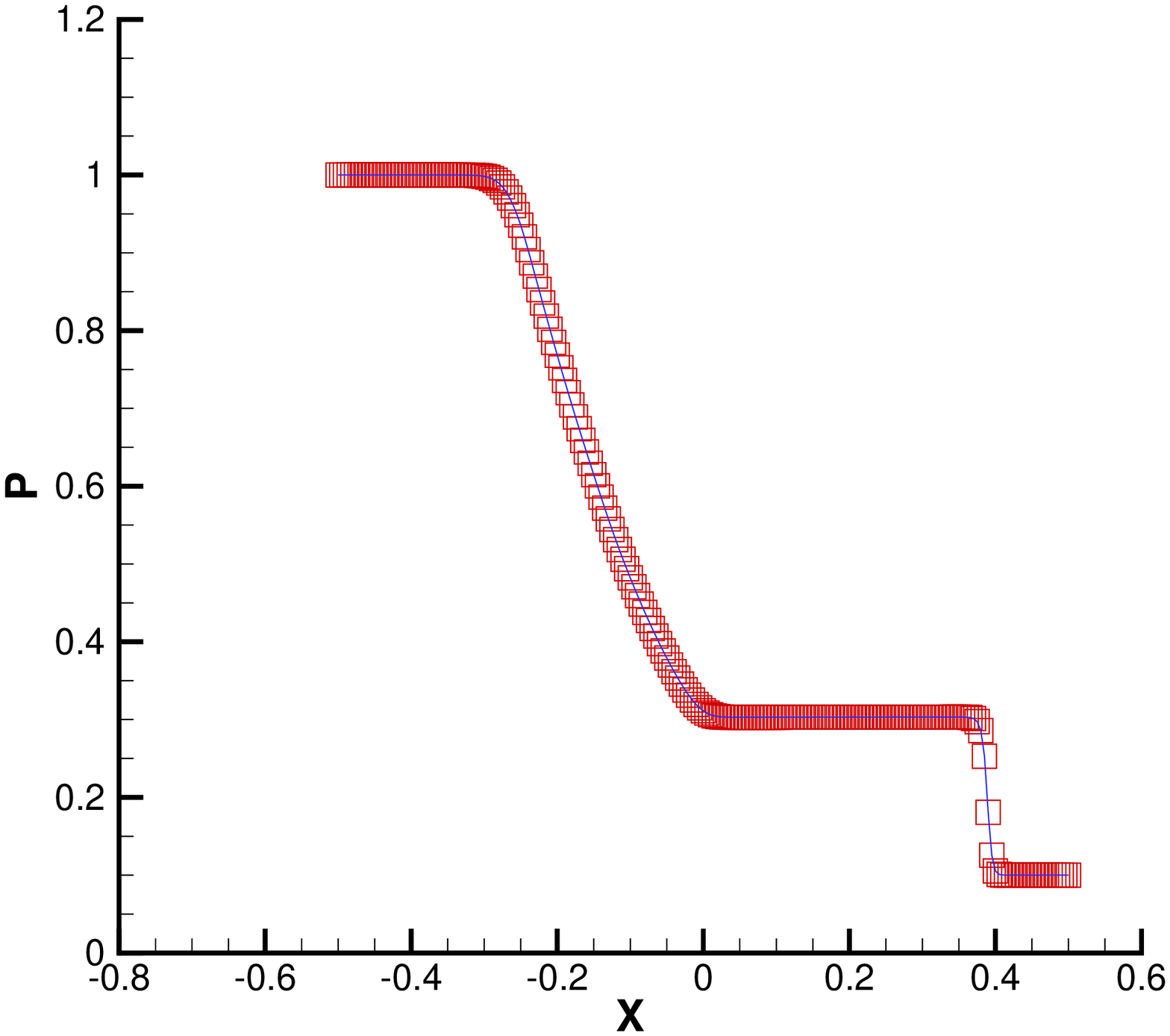}
\end{minipage}\quad
\begin{minipage}[t]{0.45\linewidth}
\centering
\includegraphics[width=\linewidth]{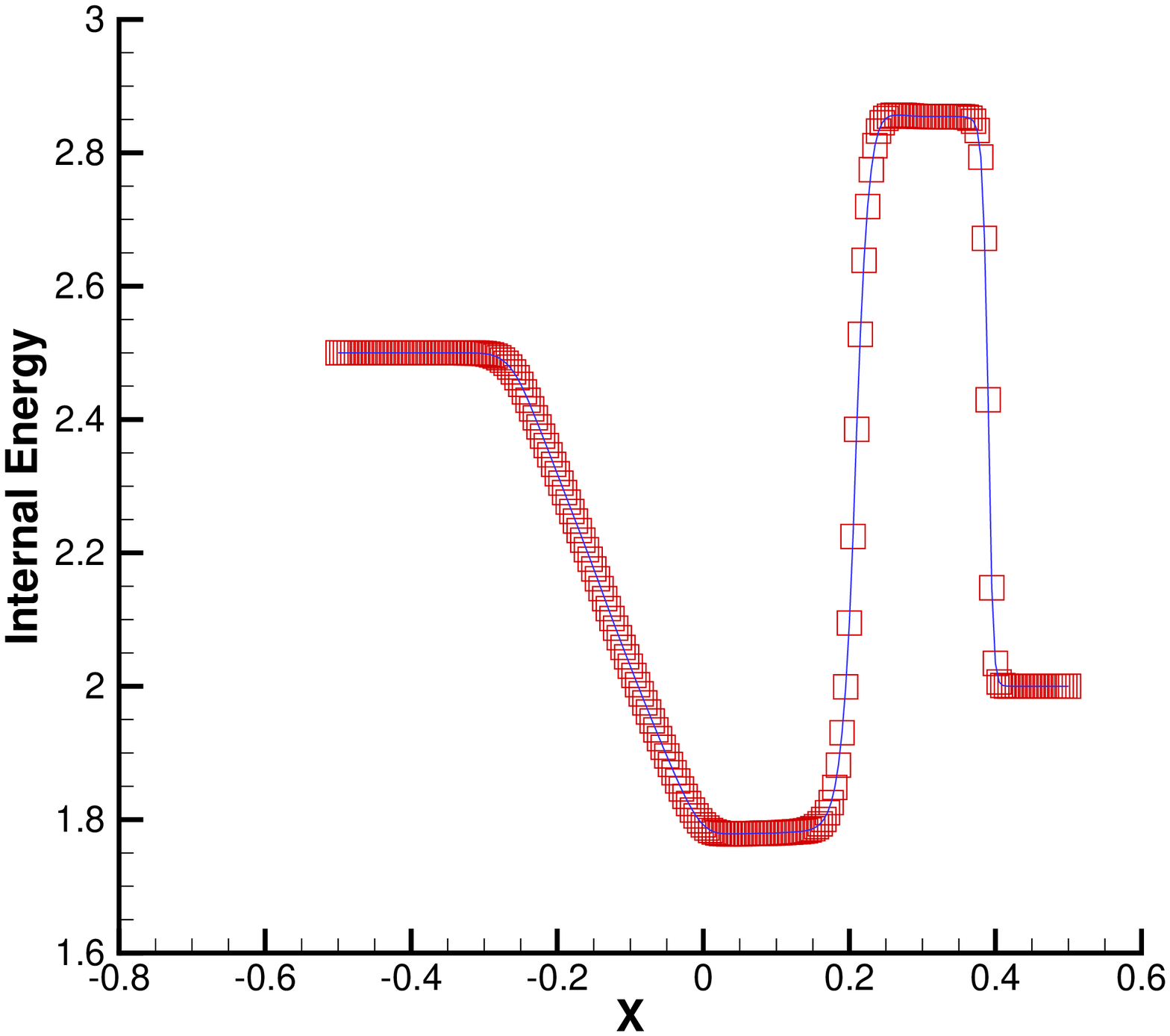}
\end{minipage}\\
\caption{Velocity (left up), density (right up), pressure (left bottom) and internal energy (right bottom) profiles of Sod shock tube.} \label{SobProblem}
\end{figure}

In this case, we set $\hat{\tau} = 10^{-4}$ and the $\hat{\zeta}_2 = 4$. The mesh size is
taken as $\Delta \hat{x} = 1/201$ and the time step size is chosen as $\Delta \hat{t} = \hat{\tau}/4$. Before the waves propagate to the two boundary ends,
the distribution functions at the boundary can be set as the
equilibrium distribution functions computed from the initial
value of macroscopic variables. Figure \ref{SobProblem} shows the computed
density, velocity, pressure, and internal energy profiles
(symbols) at $\hat{t}=0.22$, exact solutions (solid lines) are also displayed in this figure. Clearly, the present results agree
excellently well with the exact solution.

\subsection{Lax shock tube}
\begin{figure}[!h]
\centering
\begin{minipage}[t]{0.45\linewidth}
\centering
\includegraphics[width=\linewidth]{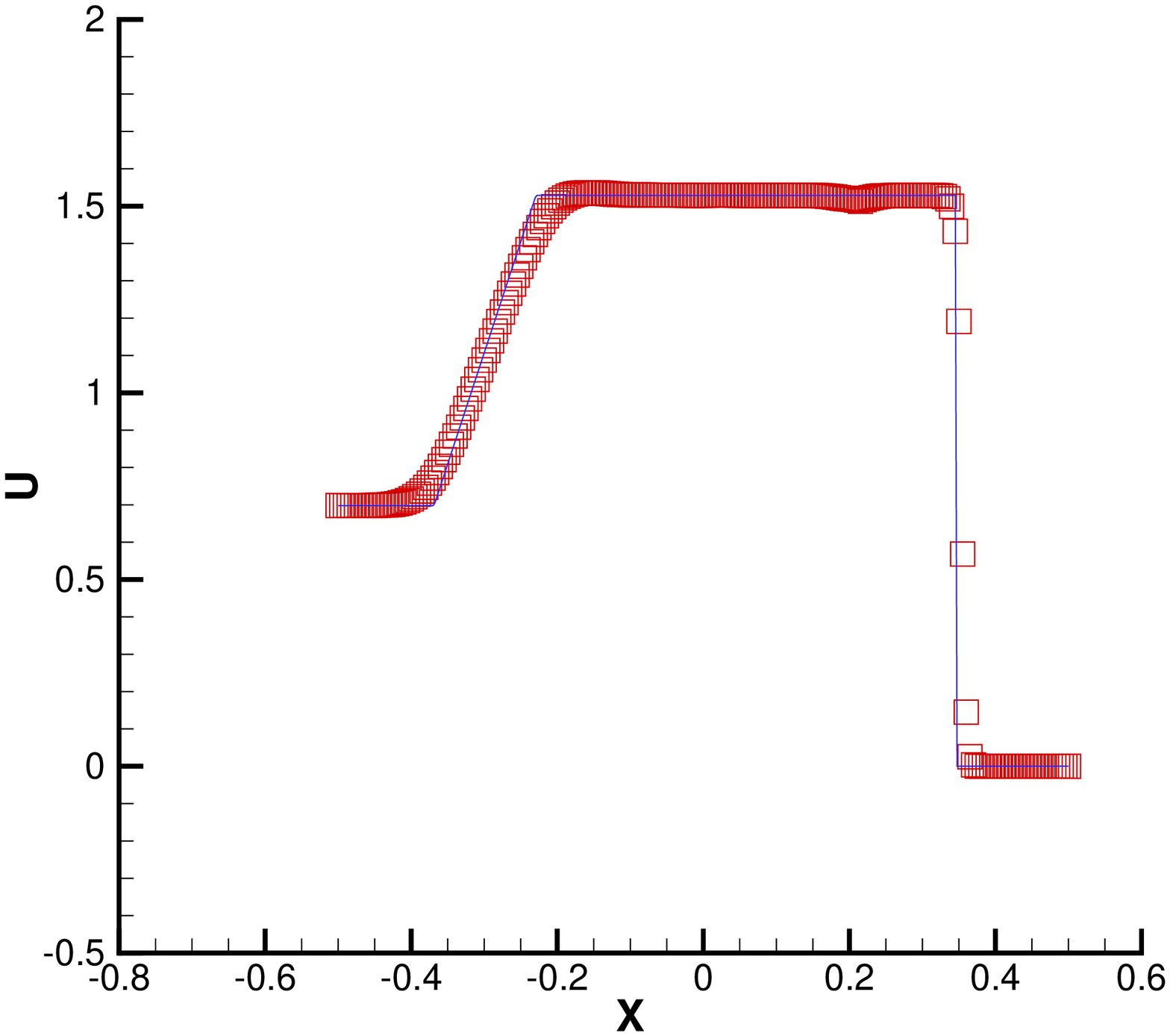}
\end{minipage}\quad
\begin{minipage}[t]{0.45\linewidth}
\centering
\includegraphics[width=\linewidth]{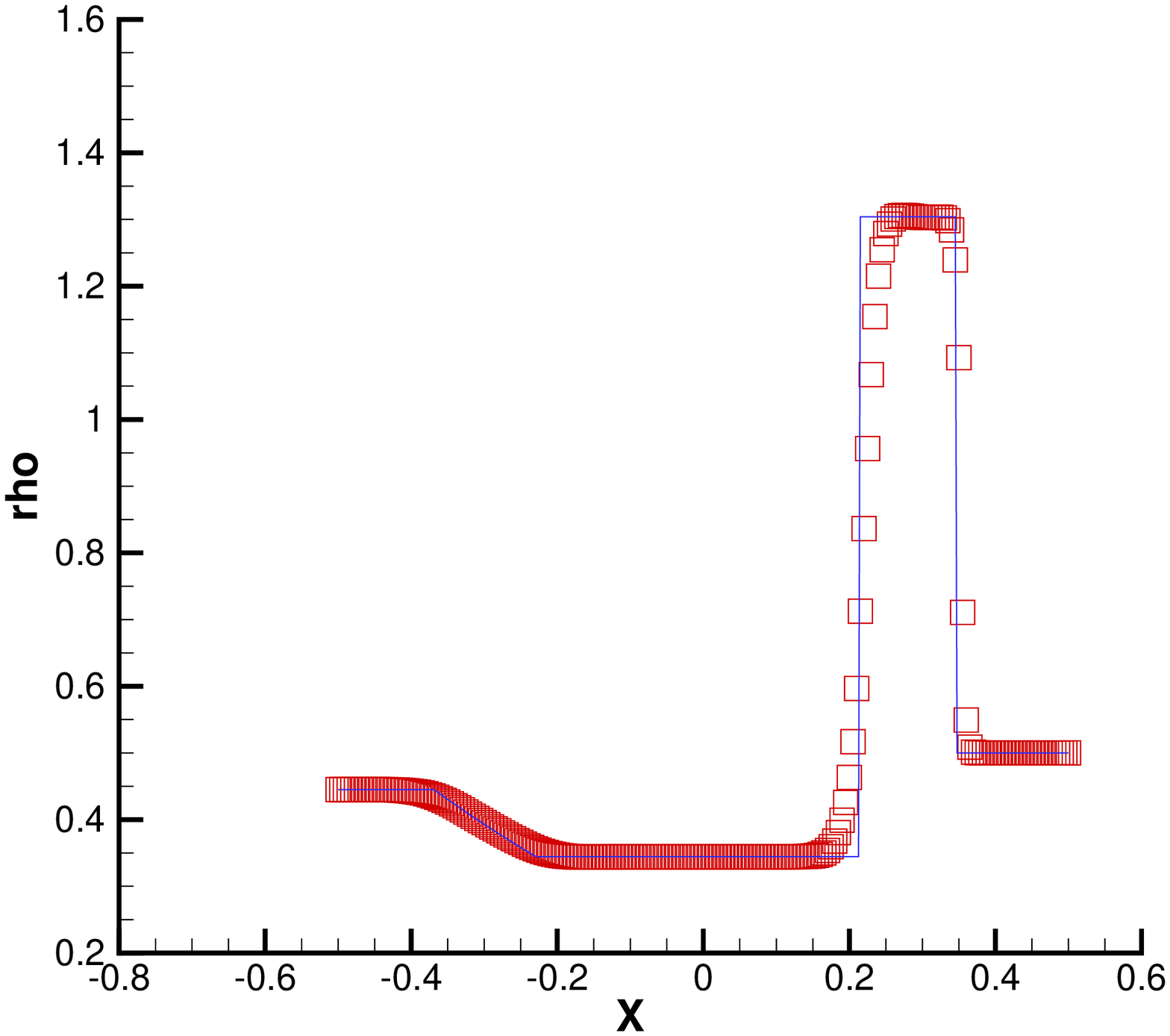}
\end{minipage}\\
\begin{minipage}[t]{0.45\linewidth}
\centering
\includegraphics[width=\linewidth]{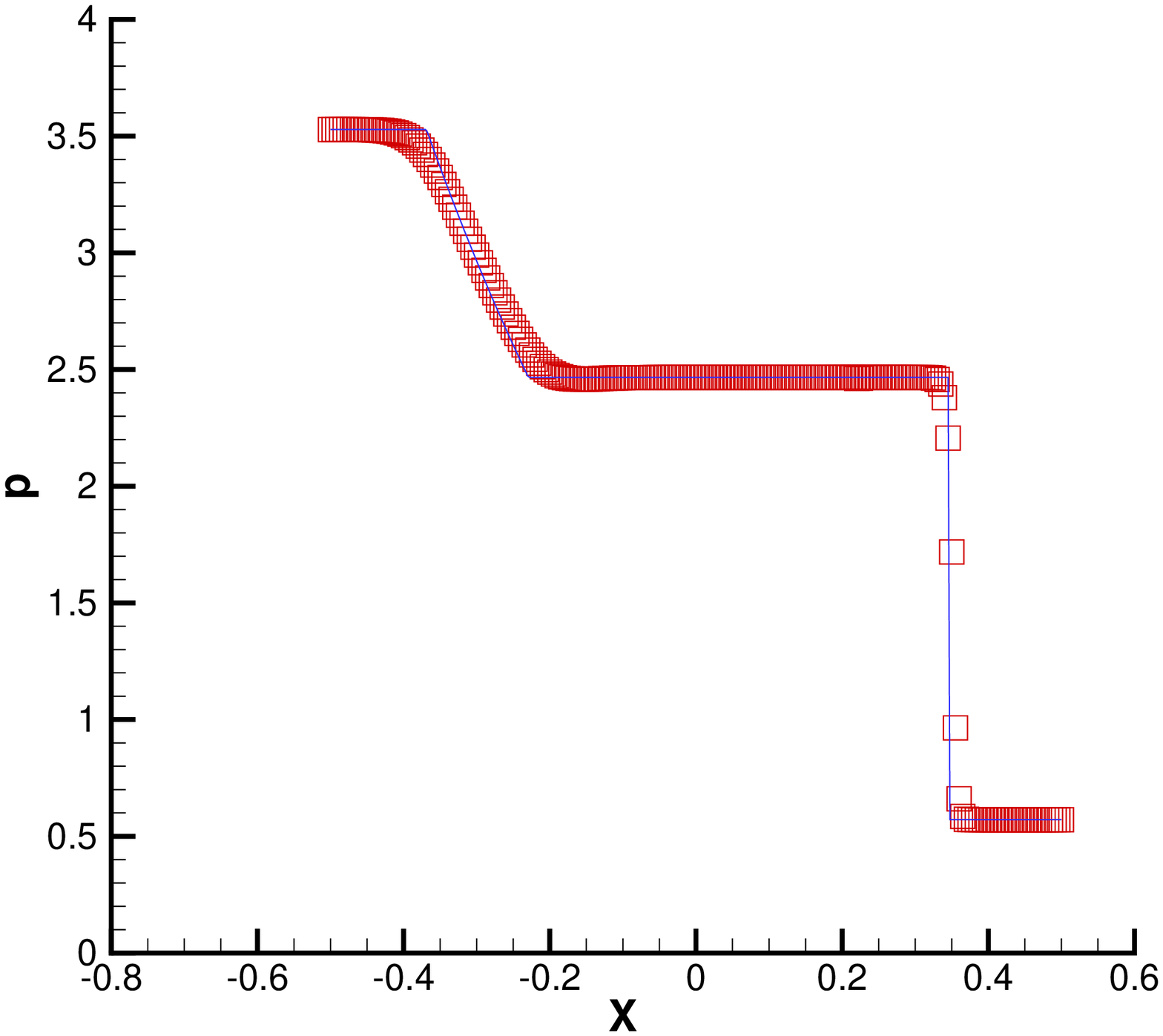}
\end{minipage}\quad
\begin{minipage}[t]{0.45\linewidth}
\centering
\includegraphics[width=\linewidth]{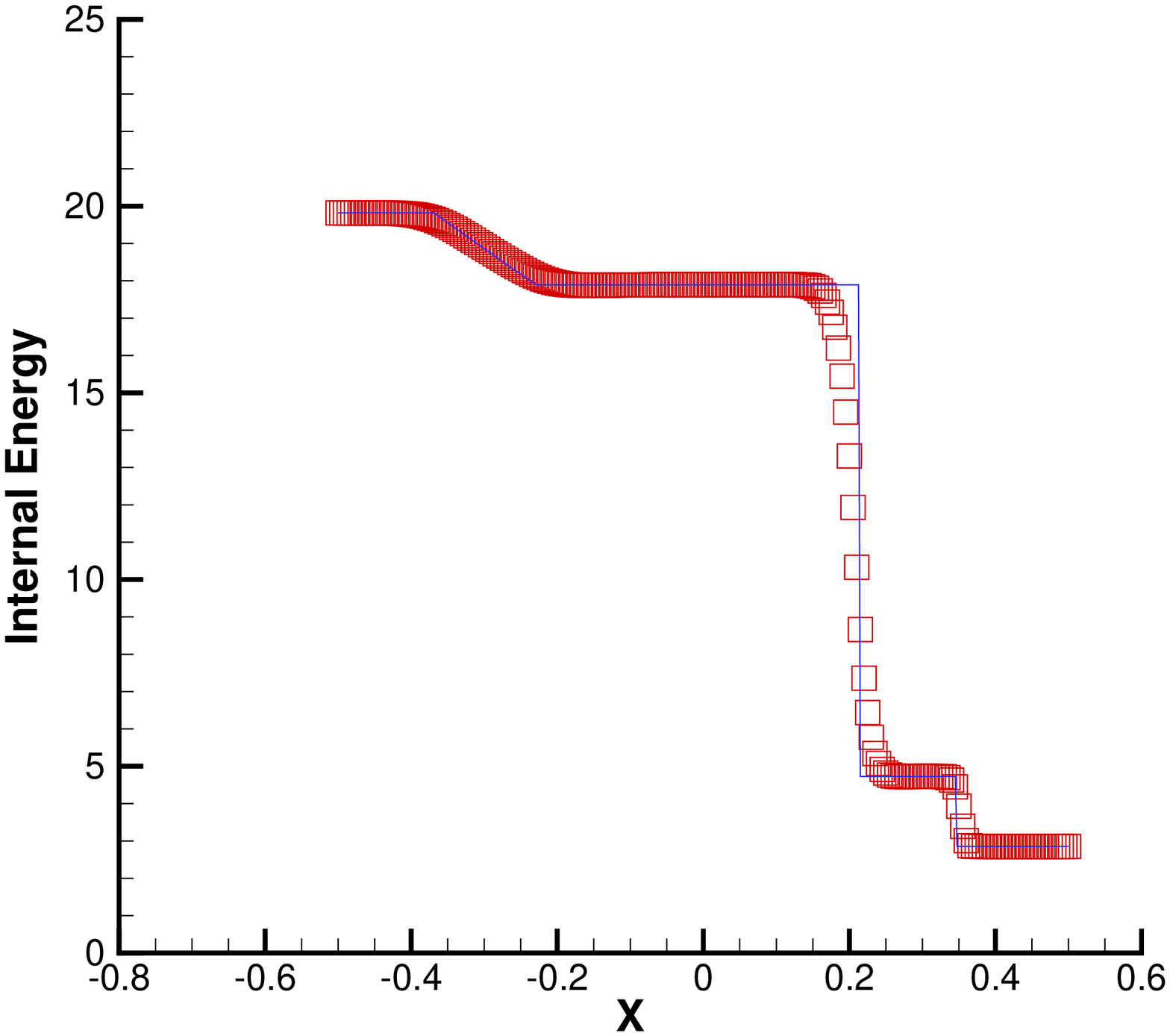}
\end{minipage}\\
\caption{Velocity (left up), density (right up), pressure (left bottom) and internal energy (right bottom) profiles of Lax shock tube.} \label{LaxProblem}
\end{figure}

The second test case is the Lax shock tube. The
initial condition of the problem is given as
\begin{equation}
\begin{aligned}
(\hat{\rho}_L,\hat{u}_L,\hat{e}_L) &= (0.445,0.698,19.82)\quad (-0.5< x < 0)\\
(\hat{\rho}_R,\hat{u}_R,\hat{e}_R) &= (0.5,0,2.855)\quad (0< x < 0.5)
\end{aligned}
\end{equation}
We set $\hat{\tau} = 10^{-4}$ and $\hat{\zeta}_2 = 30$.
The mesh size and time step size are taken to be the same as
those of the Sod shock tube problem. The computed density, velocity, pressure, and internal energy profiles (symbols) at
$\hat{t}=0.14$ are shown and compared with the exact solution
(solid lines) in Figure \ref{LaxProblem}. Obviously, the present results are very
accurate.

\section{Conclusions}
In this paper, a simple method to construct local equilibrium function for one dimensional LBM was proposed. To recover the fluid equation, we classify macroscopic conditions as non-energy statistic conditions and energy statistic conditions. For non-energy statistic conditions, we introduce assignment matrix to assign non-energy statistic conditions to obtain the local equilibrium function $f_i^{(eq)}$. For energy statistic conditions, we use fixed specific rest energy, to assign  $f_i^{(eq)}$ to different energy level that make the $f_i^{(eq)}$ satisfy energy statistic conditions. Numerical experiments showed that compressible
inviscid flows with weak and strong shock waves can be well simulated
by the present model.



%
\end{document}